\documentclass[aps,prb,twocolumn,showkeys,preprintnumbers,superscriptaddress,amsmath,amssymb]{revtex4}

\usepackage{graphicx}
\usepackage{dcolumn}
\usepackage{bm}

\begin{document}

\title{First-principles studies of the effects of impurities on the ionic and electronic conduction in LiFePO$_{4}$}

\author{Khang Hoang}
\affiliation{Center for Computational Materials Science, Naval Research Laboratory, Washington, DC 20375, USA}%
\affiliation{Computational Materials Center, George Mason University, Fairfax, Virginia 22030, USA}
\author{Michelle D. Johannes}
\email{michelle.johannes@nrl.navy.mil}
\affiliation{Center for Computational Materials Science, Naval Research Laboratory, Washington, DC 20375, USA}%

\date{\today}

\begin{abstract}

Olivine-type LiFePO$_{4}$ is widely considered as a candidate for Li-ion battery electrodes, yet its applicability in the pristine state is limited due to poor ionic and electronic conduction. Doping can be employed to enhance the material's electrical conductivity. However, this should be understood as incorporating electrically active impurities to manipulate the concentration of native point defects such as lithium vacancies and small hole polarons which are responsible for ionic and electronic conduction, respectively,  and {\it not} as generating band-like carriers. Possible effects of monovalent (Na, K, Cu, and Ag), divalent (Mg and Zn), trivalent (Al), tetravalent (Zr, C, and Si), and pentavalent (V and Nb) impurities on the ionic and electronic conductivities of LiFePO$_{4}$ are analyzed based on results from first-principles density-functional theory calculations. We identify impurities that are effective (or ineffective) at enhancing the concentration of lithium vacancies or small hole polarons. Based on our studies, we discuss specific strategies for enhancing the electrical conductivity in LiFePO$_{4}$ and provide suggestions for further experimental studies.

\end{abstract}


\maketitle

\section{\label{sec:intro}Introduction}

Olivine-type LiFePO$_{4}$ has been proposed as a candidate for rechargeable Li-ion battery electrodes because of its structural and chemical stabilities, high intercalation voltage, high theoretical discharge capacity, environmental friendliness, and potentially low costs.\cite{padhi:1188,Ellis2010,Manthiram2011} However, poor ionic and electronic conduction are major challenges. In addition to nanostructuring and carbon coating, doping has been considered as an important path toward enhancing the electrical conductivity. Since it was reported by Chung et al.~\cite{Chung:2002p246} that doping LiFePO$_{4}$ with impurities such as Mg, Ti, Zr, and Nb significantly enhances the conductivity, there have been numerous experimental works carried out along this path. Nonetheless the role of dopants in the conductivity enhancement is still under debate.\cite{Ravet2003,Wagemaker2008} Thus it is important to understand the mechanisms for ionic and electronic conduction and the effects of impurities on the ionic and electronic conductivities. Such an understanding is essential in formulating strategies for improving the electrical (i.e., ionic and/or electronic) conductivity and hence the electrochemical performance of the material.

Recently, we have carried out comprehensive first-principles density-functional theory (DFT) studies of native point defects and defect complexes in LiFePO$_{4}$.\cite{hoang2011} Based on a detailed analysis of the structure, energetics, and migration of the native defects, we arrived at the following main conclusions: (i) Native point defects such as small hole polarons ($p^{+}$), negatively charged lithium vacancies ($V_{\rm Li}^{-}$), negatively charged lithium antisites (Li$_{\rm Fe}^{-}$), and positively charged iron antisites (Fe$_{\rm Li}^{+}$) have low formation energies and hence are expected to be present in the material. The mobility of Fe$_{\rm Li}^{+}$ is low compared to $V_{\rm Li}^{-}$, supporting earlier suggestions that Fe$_{\rm Li}^{+}$ impedes Li diffusion and hence reduces the electrochemical activity.\cite{Axmann:2009p137,Chung2010} The relative concentrations of these defects are, however, sensitive to the experimental conditions during synthesis. This suggests that one can suppress or enhance certain native defects via tuning the synthesis conditions. (ii) Native defects in LiFePO$_{4}$ cannot act as sources of band-like electrons and holes, and the material cannot be doped $n$- or $p$-type. Any attempt to deliberately shift the Fermi level to the valence-band maximum (VBM) or conduction-band minimum (CBM), e.g., via doping with acceptors or donors, will result in spontaneous formation of compensating native defects that counteract the effects of doping. (iii) The ionic conduction occurs via diffusion of $V_{\rm Li}^{-}$ whereas the electronic conduction proceeds via hopping of $p^{+}$, which confirms earlier suggestions of a polaronic mechanism in the electronic conduction.\cite{Zhou:2004p101,Maxisch:2006p103,Ellis2006,Zaghib2007} The ionic conduction is effectively one-dimensional along the $b$-axis, assuming there are no other native defects or extrinsic impurities with low mobility that block the Li channels; whereas the electronic conduction is effectively two-dimensional in the $b$-$c$ plane.

It is evident from the conduction mechanisms of LiFePO$_{4}$ that the ionic and electronic conductivities are dependent on mobility and concentration, and hence on the migration barriers and formation energies, of $V_{\rm Li}^{-}$ and $p^{+}$. This opens the door to manipulating the ionic and electronic conductivities of the material via manipulating defect concentrations. In fact, LiFePO$_{4}$ samples with high concentrations of $V_{\rm Li}^{-}$ and/or $p^{+}$ (and a low concentration of Fe$_{\rm Li}^{+}$) can be obtained in experiment under appropriate synthesis conditions.\cite{hoang2011} Also, despite the fact that LiFePO$_{4}$ cannot be doped $n$- or $p$-type, one can still incorporate into the material electrically active impurities that are effective in shifting the position of the Fermi level away from that determined by the intrinsic native defects, slightly toward the CBM (for impurities that exhibit a donor-like effect) or VBM (for impurities that exhibit an acceptor-like effect). This results in lowering the formation energy and hence enhancing the concentration of $V_{\rm Li}^{-}$ or $p^{+}$. Thus ``doping'' in LiFePO$_{4}$ should be understood as a way to manipulate the concentration of the native defects that are responsible for ionic and electronic conduction, rather than as generating band-like carriers.

\begin{figure}
\begin{center}
\includegraphics[width=3.4in]{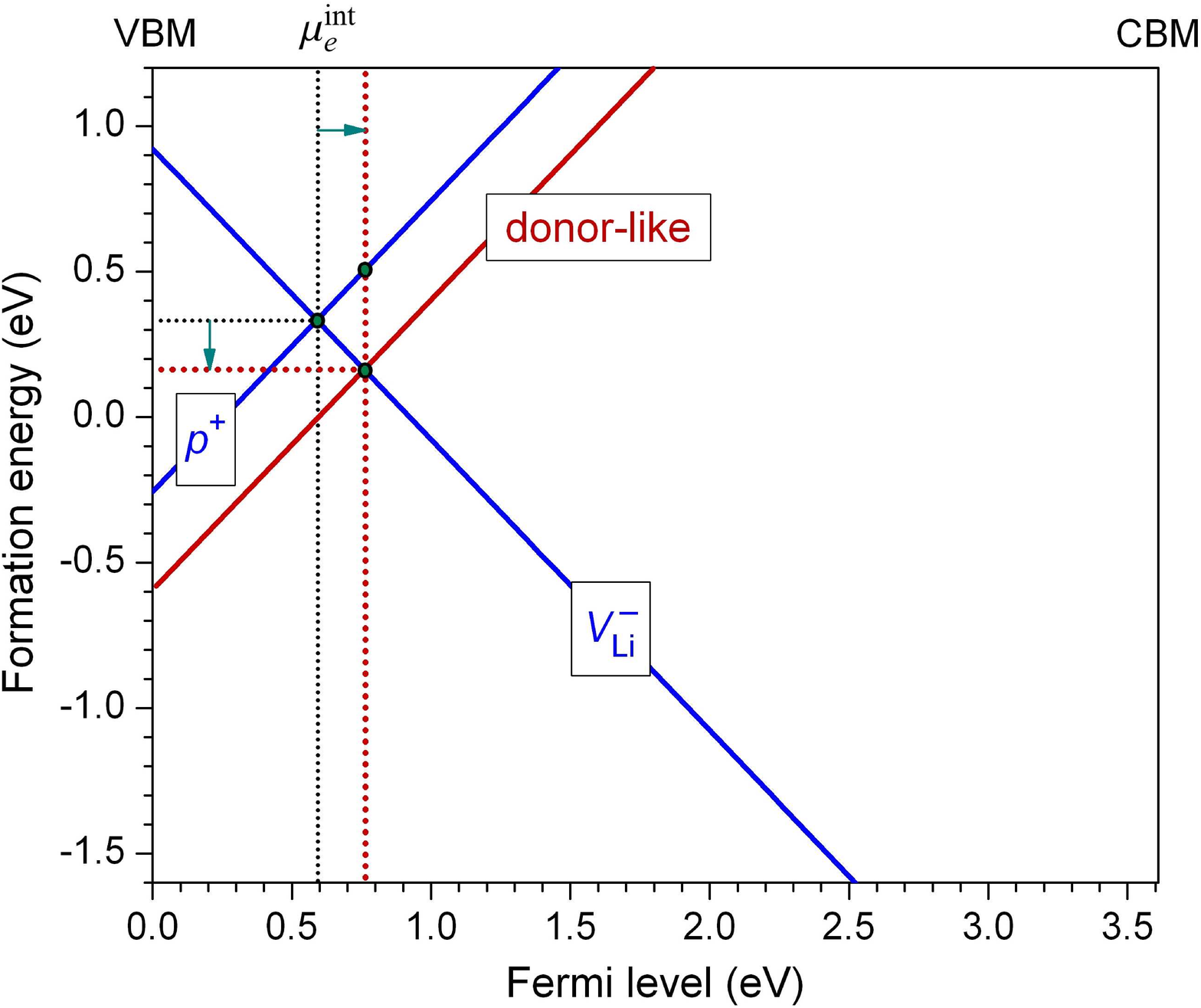}
\end{center}
\vspace{-0.2in}
\caption{Formation energies of intrinsic/native defects $V_{\rm Li}^{-}$ and $p^{+}$ (solid blue lines) and of a donor-like impurity (solid red line), plotted as a function of Fermi level with respect to the VBM. In the absence of extrinsic electrically active impurities, the Fermi level of LiFePO$_{4}$ is at $\mu_{e}^{\rm int}$ where the formation energies and thus concentrations of $V_{\rm Li}^{-}$ and $p^{+}$ are equal. When the donor-like impurity is incorporated into the material with a concentration higher than that of the positively charged native defects, the Fermi level is shifted toward the CBM, lowering (increasing) the formation energy of $V_{\rm Li}^{-}$ ($p^{+}$). As a result, the ionic (electronic) conductivity will be enhanced (reduced).}
\label{fig;Fermi}
\end{figure}

The mechanism for manipulating the concentrations of $p^{+}$ and $V_{\rm Li}^{-}$ by incorporating a donor-like impurity is illustrated in Fig.~\ref{fig;Fermi}. Note that, in the absence of extrinsic electrically active impurities, the Fermi level of insulating materials such as LiFePO$_{4}$ is at $\mu_{e}$=$\mu_{e}^{\rm int}$, determined by a charge neutrality condition that involves all possible intrinsic/native defects. This condition requires that defects with different charge states coexist in a proportion that maintains overall charge neutrality in the material.\cite{peles2007, hoang2009, wilson-short, hoang_amide_angew} Because of the exponential dependence of the concentration on defect formation energy (see the next section), $\mu_{e}^{\rm int}$ is predominantly determined by the positively and negatively charged defects with the lowest formation energies (In the illustrated example, these defects are $V_{\rm Li}^{-}$ and $p^{+}$). If electrically active impurities are incorporated into LiFePO$_{4}$ with concentrations higher than that of the native defects, the Fermi level may be shifted away from $\mu_{e}^{\rm int}$ as the charge neutrality condition is re-established,\cite{peles2007, hoang2009} as illustrated in Fig.~\ref{fig;Fermi}.

Clearly, it is important to know which impurities are effective at enhancing which native defect(s). In this Article, we report our first-principles studies of various monovalent (Na, K, Cu, and Ag), divalent (Mg and Zn), trivalent (Al), tetravalent (Zr, C, and Si), and pentavalent (V and Nb) impurities in LiFePO$_{4}$ and possible defect complexes between the extrinsic impurities and native defects. The effects of the impurities on the ionic and electronic conduction will be discussed and comparisons with available experimental data will be made where appropriate. On the basis of our studies, we discuss specific strategies for enhancing the electrical conductivity in LiFePO$_{4}$.

\section{\label{sec;metho}Methodology}

Our first-principles calculations of native defects and extrinsic impurities (hereafter commonly referred to as ``defects'') were based on density-functional theory within the GGA+$U$ framework,\cite{anisimov1991,anisimov1993,liechtenstein1995} which is an extension of the generalized-gradient approximation,\cite{GGA} and the projector augmented wave method,\cite{PAW1,PAW2} as implemented in the VASP code.\cite{VASP1,VASP2,VASP3} The $U$ value for Fe was taken from Zhou et al.~\cite{Zhou:2004p104}. We used an orthorhombic (1$\times$2$\times$2) supercell, corresponding to 112 atoms per cell, to model the defects in LiFePO$_{4}$.

Different defects in LiFePO$_{4}$ are characterized by their formation energies. The formation energy ($E^{f}$) of a defect is a crucial factor in determining its concentration. In thermal equilibrium, the concentration of the defect X at temperature $T$ can be obtained via the relation\cite{hoang2011, walle:3851} 
\begin{equation}\label{eq;concen} 
c(\mathrm{X})=N_{\mathrm{sites}}N_{\mathrm{config}}\mathrm{exp}[-E^{f}(\mathrm{X})/k_BT], 
\end{equation} 
where $N_{\mathrm{sites}}$ is the number of high-symmetry sites in the lattice per unit volume on which the defect can be incorporated, and $N_{\mathrm{config}}$ is the number of equivalent configurations (per site). It emerges from Eq.~(\ref{eq;concen}) that defects with lower formation energies will occur in higher concentrations.

The formation energy of a defect X in charge state $q$ is defined as\cite{hoang2011, walle:3851} 
\begin{eqnarray}\label{eq;eform}
\nonumber
E^{f}({\mathrm{X}}^q)=E_{\mathrm{tot}}({\mathrm{X}}^q)-E_{\mathrm{tot}}({\mathrm{bulk}})-\sum_{i}{n_i\mu_i}\\
+q(E_{\mathrm{v}}+\Delta V+\mu_{e}), 
\end{eqnarray} 
where $E_{\mathrm{tot}}(\mathrm{X}^{q})$ and $E_{\mathrm{tot}}(\mathrm{bulk})$ are the total energies of a supercell containing X and of a supercell of the perfect bulk material; $\mu_{i}$ is the atomic chemical potential of species $i$ (and is referenced to the standard state), and $n_{i}$ denotes the number of atoms of species $i$ that have been added ($n_{i}$$>$0) or removed ($n_{i}$$<$0) to form the defect. $\mu_{e}$ is the electron chemical potential (the Fermi level) referenced to the VBM in the bulk ($E_{\mathrm{v}}$). $\Delta V$ is the ``potential alignment'' term, i.e., the shift in the band positions due to the presence of the charged defect and the neutralizing background, obtained by aligning the average electrostatic potential in regions far away from the defect to the bulk value.\cite{walle:3851}

The atomic chemical potentials $\mu_{i}$ are variables and can be chosen to represent experimental conditions, and are subject to various thermodynamic constraints.\cite{hoang2011} These constraints ensure that LiFePO$_{4}$ is thermodynamically stable. The calculated phase diagrams of the quaternary Li-Fe-P-O$_{2}$ system at 0 K reported by Ong et al.~\cite{ong2008} show that the compound is stable over a range of the oxygen chemical potential ($\mu_{{\rm O}_{2}}$) values, from $-$3.03 to $-$8.25 eV. These two values are the upper and lower limits in the range of $\mu_{{\rm O}_{2}}$ values considered in our work. Lower $\mu_{{\rm O}_{2}}$ values represent the so-called ``more reducing environments,'' which are usually associated with higher temperatures and lower oxygen partial pressures and/or the presence of oxygen reducing agents; whereas higher $\mu_{{\rm O}_{2}}$ values represent ``less reducing environments.''\cite{ong2008} For each $\mu_{{\rm O}_{2}}$, the chemical potential of Li, Fe, and P can be defined through a set of constraints relevant at that $\mu_{{\rm O}_{2}}$ value. For example, at $\mu_{{\rm O}_{2}}$=$-$4.59 eV, the thermodynamically allowed range of $\mu_{{\rm Li}}$ and $\mu_{{\rm Fe}}$ values is the area defined by secondary phases Fe$_{2}$O$_{3}$, Fe$_{3}$(PO$_{4}$)$_{2}$, Fe$_{2}$P$_{2}$O$_{7}$, Li$_{4}$P$_{2}$O$_{7}$, and Li$_{3}$PO$_{4}$ in the chemical-potential diagram as depicted in Fig.~1 of Ref.~\cite{hoang2011}; the remaining variable $\mu_{{\rm P}}$ is determined via the condition that ensures the stability of LiFePO$_{4}$.\cite{hoang2011} Our investigation of thermodynamically allowed $\mu_{{\rm Li}}$, $\mu_{{\rm Fe}}$, $\mu_{{\rm P}}$, and $\mu_{{\rm O}_{2}}$ values thus involves all possible Li-Fe-P-O$_{2}$ phases reported in Ref.~\cite{ong2008}. Environments under which the system is close to forming Li-containing (Fe-containing) secondary phases are referred to as Li-excess (Li-deficient).\cite{hoang2011, ong2008}

For the extrinsic impurities, the lower limit on $\mu_{i}$ is minus infinity and the upper limit is zero (with respect to the energy of the elemental bulk phase).\cite{walle:3851} Although stronger bounds on the impurity chemical potentials can be estimated based on other solubility-limiting phases formed between the impurities and the host constituents, there is no way to know their exact values. In the following presentation, the atomic chemical potentials of the impurities were chosen somewhat arbitrarily within the range given by the above mentioned upper and lower limits, provided that the calculated formation energies at $\mu_{e}^{\rm int}$ are positive. These choices, however, in no way affect the physics of what we are presenting since we are interested only in the relative formation energies of the impurities at a given set of Li, Fe, P, and O$_{2}$ chemical potentials. In other words, we are trying to answer the following question: Once a certain impurity is incorporated into the material, what is its lattice site preference under thermodynamic equilibrium conditions and what is its possible effect on the material's properties? Further details of the calculations, our methodology (and its limitations), and the results for native defects in LiFePO$_{4}$ can be found in Ref.~\cite{hoang2011}.

\section{Results and discussion}

Impurities in LiFePO$_{4}$ can be substitutional at the Li, Fe, and P sites, or may be at interstitial sites. For each substitutional or interstitial impurity, calculations were carried out in several possible charge states. We paid attention to positively and negatively charged impurities that have the lowest formation energy at $\mu_{e}^{\rm int}$, and hence the highest concentration, since they can act as acceptor-like (donor-like) dopants that shift the Fermi level toward the VBM (CBM). Note that $\mu_{e}^{\rm int}$ is dependent on the chemical potentials. For example, we find that $\mu_{e}^{\rm int}$=0.58$-$0.59 eV for $\mu_{{\rm O}_{2}}$=$-$3.03 eV, 0.72$-$1.06 eV for $\mu_{{\rm O}_{2}}$=$-$4.59 eV, and 1.98$-$2.00 eV for $\mu_{{\rm O}_{2}}$=$-$8.21 eV.\cite{hoang2011} We also considered possible complexes consisting of extrinsic impurities and native point defects. In the following, we present only results for those defect configurations that have low formation energies and hence are potentially relevant to the material's properties. Note that the defect formation energies presented in all the figures were evaluated at $\mu_{{\rm O}_{2}}$=$-$3.03 eV (corresponding to, e.g., a temperature of 800$^{\circ}$C and an oxygen partial pressure of 10$^{-3}$ atm, Ref.~\cite{Reuter2001}), and equilibrium between LiFePO$_{4}$, Fe$_{2}$O$_{3}$, and Fe$_{7}$(PO$_{4}$)$_{6}$ was assumed (i.e., Li-deficient environment).\cite{hoang2011} These conditions give rise to $\mu_{{\rm Li}}$=$-$3.41 eV, $\mu_{{\rm Fe}}$=$-$3.35 eV, and $\mu_{{\rm P}}$=$-$6.03 eV with the formation enthalpies of the Li-Fe-P-O$_{2}$ phases taken from Ref.~\cite{ong2008}. Under these conditions, $p^{+}$ and $V_{\rm Li}^{-}$ are the dominant native point defects and the Fermi level is at $\mu_{e}^{\rm int}$=0.59 eV,\cite{hoang2011} as illustrated in Fig.~\ref{fig;Fermi}. Of course the energy landscape may be different for a different set of Li, Fe, P, and O$_{2}$ chemical potentials. Therefore, we will extend our discussions and conclusions to all thermodynamically allowed chemical potentials mentioned in the previous section.

\subsection{Monovalent impurities}

\begin{figure}
\begin{center}
\includegraphics[width=3.4in]{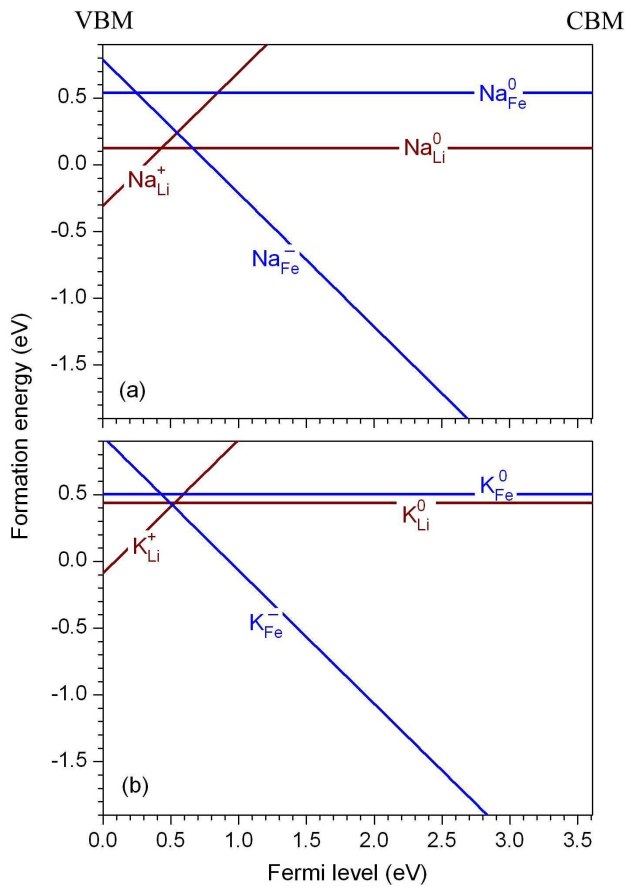}
\end{center}
\vspace{-0.2in}
\caption{Formation energies of (a) Na- and (b) K-related impurities at $\mu_{{\rm O}_{2}}$=$-$3.03 eV, plotted as a function of Fermi level with respect to the VBM.}\label{fig;Na-K}
\end{figure}

Figures \ref{fig;Na-K}(a) and \ref{fig;Na-K}(b) show the formation energies of Na- and K-related substitutional impurities at the Li and Fe sites, evaluated at $\mu_{{\rm O}_{2}}$=$-$3.03 eV and its associated $\mu_{{\rm Li}}$, $\mu_{{\rm Fe}}$, and $\mu_{{\rm P}}$ values as noted earlier. The slope in the formation energy plots indicates the charge state. Positive slope indicates that the defect is positively charged, negative slope indicates the defect is negatively charged; cf.~Eq~(\ref{eq;eform}). We find that Na$_{\rm Li}^{0}$ and Na$_{\rm Fe}^{-}$ are elementary point defects; whereas Na$_{\rm Li}^{+}$ is, in fact, a complex of Na$_{\rm Li}^{0}$ and small hole polaron $p^{+}$ that is stabilized at the neighboring Fe site, and Na$_{\rm Fe}^{0}$ a complex of Na$_{\rm Fe}^{-}$ and $p^{+}$. At the Fe site where the hole polaron is located, the average Fe$-$O bond length is 2.06 {\AA} (compared to 2.18 {\AA} of the other Fe-O bonds) and the calculated magnetic moment is 4.28 $\mu_{\rm B}$ (compared to 3.76 $\mu_{\rm B}$ at other Fe sites); these numbers are in excellent agreement with those obtained for an isolated hole polaron in LiFePO$_{4}$.\cite{hoang2011} For all possible Li, Fe, P, and O$_{2}$ chemical potentials, Na$_{\rm Li}^{0}$ has the lowest formation energy at $\mu_{e}^{\rm int}$, as seen in Fig.~\ref{fig;Na-K}(a) for $\mu_{{\rm O}_{2}}$=$-$3.03 eV, suggesting that Na is likely to exist as neutral Na$_{\rm Li}^{0}$ under thermodynamic equilibrium and therefore not effective in enhancing either $p^{+}$ or $V_{\rm Li}^{-}$. The formation energy difference between Na$_{\rm Li}^{0}$ and Na$_{\rm Fe}^{-}$ is, however, small (about 0.25 eV at most, as measured at $\mu_{e}^{\rm int}$).

Strictly speaking, Na$_{\rm Li}^{+}$ is not the positive charge state of substitutional Na at the Li site, and Na$_{\rm Fe}^{0}$ is not the neutral charge state of Na at the Fe site. In other words, Na$_{\rm Li}$, as a single point defect, is only stable as Na$_{\rm Li}^{0}$, and Na$_{\rm Fe}$ as Na$_{\rm Fe}^{-}$, since locally stable configurations of charge states other than Na$_{\rm Li}^{0}$ and Na$_{\rm Fe}^{-}$ cannot be stabilized. If we try, e.g., to create Na$_{\rm Li}^{+}$, it decays to a situation where the positive charge is not associated with the point defect but corresponds to a small hole polaron stabilized at the neighboring Fe site, as presented above. Note that, instead of writing down, e.g., Na$_{\rm Li}^{+}$ as Na$_{\rm Li}^{0}$-$p^{+}$, we keep the notation Na$_{\rm Li}^{+}$ for simplicity, but it should be regarded as a nominal notation referring to a defect complex that consists of the stable/elementary Na$_{\rm Li}^{0}$ and a small hole polaron.

For K-related impurities, K$_{\rm Li}^{0}$ and K$_{\rm Fe}^{-}$ are elementary defects; whereas K$_{\rm Li}^{+}$ is a complex of K$_{\rm Li}^{0}$ and $p^{+}$, and K$_{\rm Fe}^{0}$ a complex of K$_{\rm Fe}^{-}$ and $p^{+}$, all similar to Na-related impurities. However, we find that, for $\mu_{{\rm O}_{2}}$=$-$3.25 eV and Li-excess or for $-$3.25 eV$<$$\mu_{{\rm O}_{2}}$$\leq$$-$3.03 eV, K$_{\rm Fe}^{-}$ and K$_{\rm Li}^{+}$ have the lowest energies for the entire range of the Fermi-level values, similar to the energy landscape presented in Fig.~\ref{fig;Na-K}(b); whereas K$_{\rm Li}^{0}$ (or K$_{\rm Fe}^{0}$, depending on the specific set of the chemical potentials) has the lowest energy for $\mu_{{\rm O}_{2}}$=$-$3.25 eV and Li-deficiency or for $-$8.21 eV$\leq$$\mu_{{\rm O}_{2}}$$<$$-$3.25 eV. The fact that K$_{\rm Fe}^{-}$ can have lower formation energy than any other K-related impurity at $\mu_{e}^{\rm int}$ suggests that this defect can shift the Fermi level toward the VBM and lower the formation energy of positively charged native defects. As a result, there will be an increase in the concentration of $p^{+}$ and hence in the electronic conductivity. It is, however, noted that the formation energy difference between K$_{\rm Fe}^{-}$ and K$_{\rm Li}^{0}$ (or K$_{\rm Fe}^{0}$) is small (about 0.15 eV at most). The difference between the general results for Na and those for K, albeit small, can be attributed to the difference in their ionic radii. Note that K can be incorporated at the Fe site only for a small range of the oxygen chemical potential ($-$3.25 eV$\leq$$\mu_{{\rm O}_{2}}$$\leq$$-$3.03 eV) under thermodynamic equilibrium, and the Fermi-level shift is relatively small (about 0.18 eV at most). The effects of K, therefore, may not be very significant.

Similar calculations were also carried out for Cu- and Ag-related impurities. We find that Cu$_{\rm Li}^{0}$ has the lowest formation energy for $\mu_{{\rm O}_{2}}$$<$$-$4.59 eV, whereas Cu$_{\rm Fe}^{0}$ has the lowest energy for $\mu_{{\rm O}_{2}}$$\geq$$-$4.59 eV. For Ag-related impurities, Ag$_{\rm Li}^{0}$ has the lowest formation energy at $\mu_{e}^{\rm int}$ for all thermodynamically allowed Li, Fe, P, and O$_{2}$ chemical potentials. Therefore, Cu and Ag are not likely to be effective in shifting the Fermi level, similar to Na.

Experimentally, it has been suggested that Na- and K-doped LiFePO$_{4}$/C samples have an improved electrochemical performance compared to the undoped LiFePO$_{4}$/C.
\cite{Yin20104308,fang_kdoped} For example, Na doping was reported to slightly enhance the electrical conductivity: 1.9$\times$10$^{-2}$ and 0.55$\times$10$^{-2}$ S.cm$^{-1}$ for, respectively, Li$_{0.97}$Na$_{0.03}$FePO$_{4}$/C and undoped LiFePO$_{4}$/C at room temperature.\cite{Yin20104308} As discussed above, Na at the Li site is electrically inactive and thus cannot enhance the concentration of $p^{+}$ or $V_{\rm Li}^{-}$ and hence the conductivity. Furthermore, Na is less likely to be incorporated at the Fe site. One, however, cannot exclude the possibility of also having a significant concentration of Na at the Fe sites, especially if the samples are prepared under conditions far from equilibrium. Given the relatively small formation energy difference between the Li and Fe sites, further experimental studies should be carried out to clarify the lattice site preference of the monovalent impurities in LiFePO$_{4}$ and its dependence on the synthesis conditions.

\subsection{Divalent impurities}

\begin{figure}
\begin{center}
\includegraphics[width=3.4in]{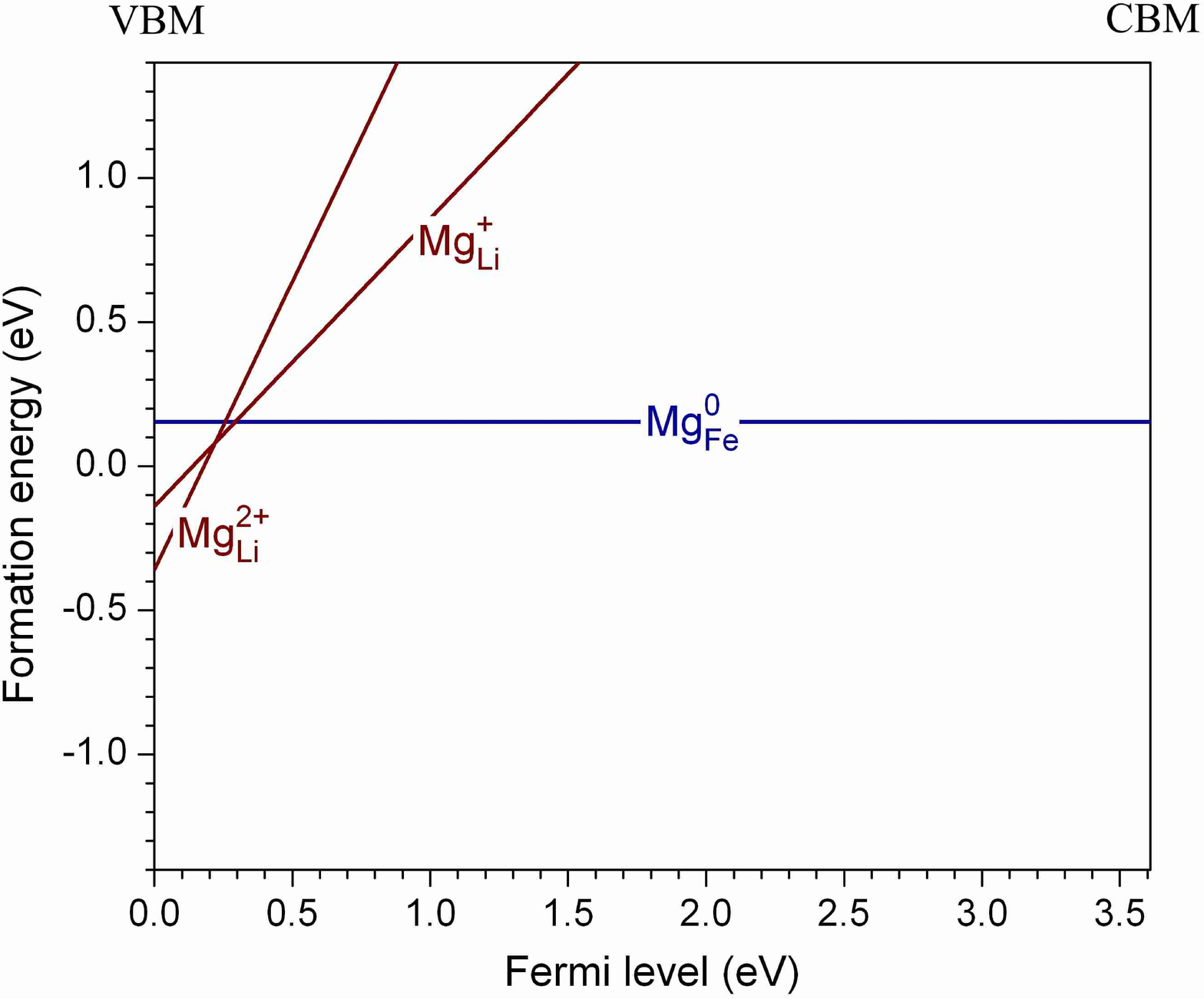}
\end{center}
\vspace{-0.2in}
\caption{Formation energies of Mg-related impurities at $\mu_{{\rm O}_{2}}$=$-$3.03 eV, plotted as a function of Fermi level with respect to the VBM.}\label{fig;Mg}
\end{figure}

Figure \ref{fig;Mg} shows the formation energies of Mg-related substitutional impurities at the Li and Fe sites, evaluated at $\mu_{{\rm O}_{2}}$=$-$3.03 eV. We find that Mg$_{\rm Fe}^{0}$ and Mg$_{\rm Li}^{+}$ are elementary point defects, whereas Mg$_{\rm Li}^{2+}$ is a complex of Mg$_{\rm Li}^{+}$ and $p^{+}$. Among all possible Mg-related impurities, Mg$_{\rm Fe}^{0}$ has the lowest formation energy at $\mu_{e}^{\rm int}$ for all possible Li, Fe, P, and O$_{2}$ chemical potentials. Note that the formation energy at $\mu_{e}^{\rm int}$ of Mg$_{\rm Li}^{+}$ is higher than that of Mg$_{\rm Fe}^{0}$ by 0.13$-$0.36 eV, depending on the specific set of the chemical potentials. In fact, the energy of Mg$_{\rm Fe}^{0}$ is lowest for almost the entire range of the Fermi-level values except very near the VBM, as seen in Fig.~\ref{fig;Mg} for $\mu_{{\rm O}_{2}}$=$-$3.03 eV. Thus Mg is likely to be electrically inactive. Zn gives similar results (not shown in the figure), i.e., Zn is likely to exist as neutral Zn$_{\rm Fe}^{0}$ and does not shift $\mu_{e}$ away from $\mu_{e}^{\rm int}$.

Experimentally, Roberts et al.~\cite{Roberts2008754} reported that there was no evidence of Mg on the Li site in samples prepared with the nominal stoichiometry Li$_{1-x}$Mg$_{0.5x}$FePO$_{4}$, which is consistent with our conclusion that Mg is likely to exist as Mg$_{\rm Fe}^{0}$ in LiFePO$_{4}$. Note that these results are in contrast to those reported by other research groups who suggested that Mg occurs on the Li site.\cite{Chung:2002p246,Guo2005113,meethong2009} Since Mg$_{\rm Fe}^{0}$ can enhance neither the concentration of $p^{+}$ nor that of $V_{\rm Li}^{-}$, the reported enhancement in the electrical conductivity of Mg-doped LiFePO$_{4}$ (Refs.~\cite{Chung:2002p246,Roberts2008754,wang:A65}) is difficult to understand as arising directly from Mg doping at the Fe site. However, given the relatively small formation energy difference at $\mu_{e}^{\rm int}$ between the Li and Fe sites (0.13$-$0.36 eV), a significant concentration of Mg at the Li sites might still be possible, especially if the samples are prepared under conditions far from equilibrium. Regarding Zn, Bilecka et al.~\cite{C0JM03476B} reported a conductivity of 10.42$\times$10$^{-8}$ S.cm$^{-1}$ at room temperature in LiZn$_{x}$Fe$_{1-x}$PO$_{4}$ samples, compared to 7.5$\times$10$^{-8}$ S.cm$^{-1}$ of undoped LiFePO$_{4}$. Since Zn$_{\rm Fe}^{0}$ is not effective in shifting the Fermi level, this small conductivity increase may be due to other effects.

\subsection{Trivalent impurities}

\begin{figure}
\begin{center}
\includegraphics[width=3.4in]{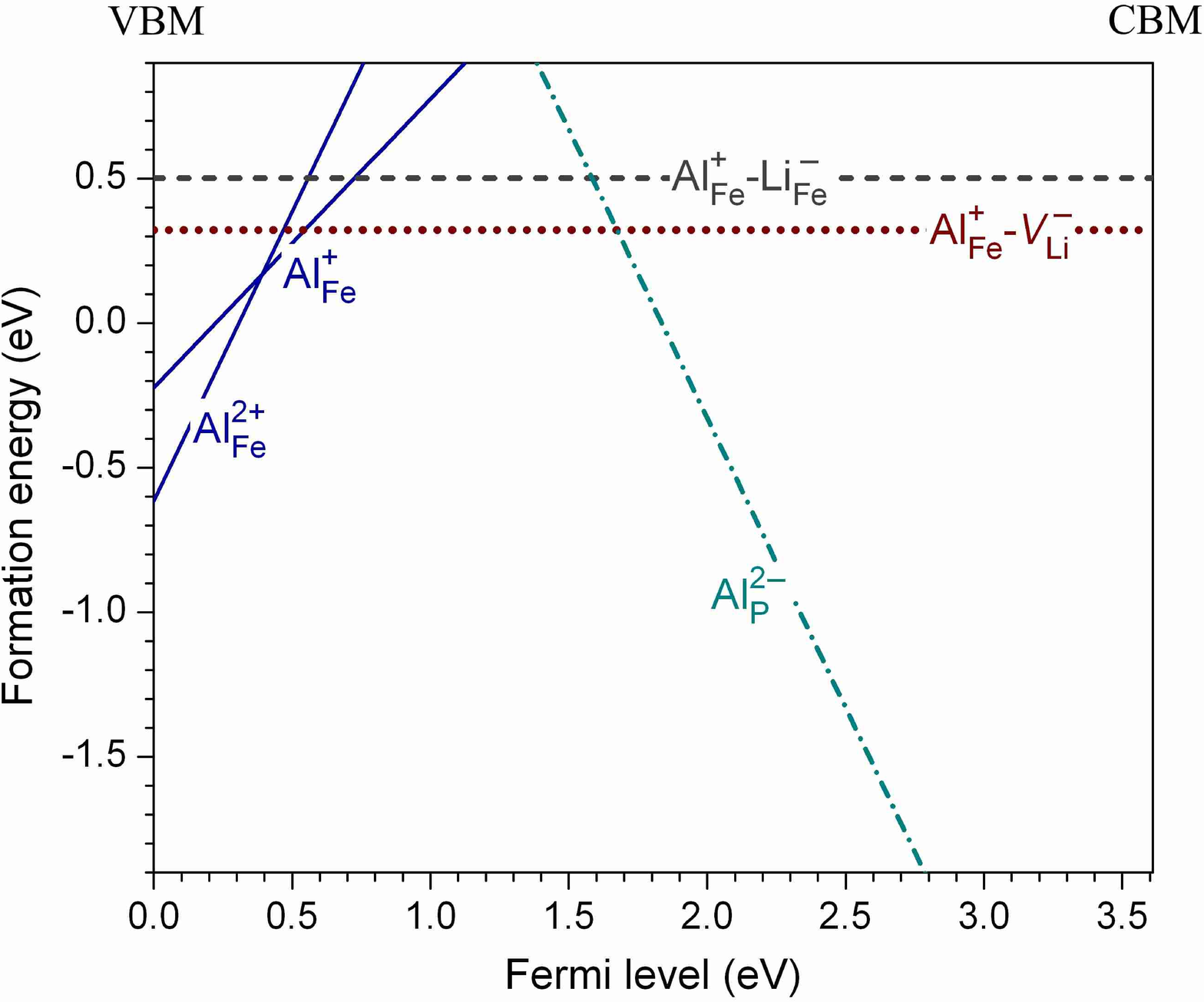}
\end{center}
\vspace{-0.2in}
\caption{Formation energies of Al-related impurities at $\mu_{{\rm O}_{2}}$=$-$3.03 eV, plotted as a function of Fermi level with respect to the VBM.}\label{fig;Al}
\end{figure} 

Figure \ref{fig;Al} shows the formation energies of Al-related substitutional impurities at the Fe and P sites, evaluated at $\mu_{{\rm O}_{2}}$=$-$3.03 eV. We find that Al$_{\rm Fe}^{+}$ is an elementary point defect, whereas Al$_{\rm Fe}^{2+}$ is a complex of Al$_{\rm Fe}^{+}$ and $p^{+}$. At the P site, Al$_{\rm P}^{2-}$ is found to have the lowest energy, which is not unexpected, given the valence of Al (+3) and P (+5). In this configuration, Al replaces P and forms a slightly distorted AlO$_{4}$ unit with the average Al$-$O bond length of 1.77 {\AA}, compared to 1.56 {\AA} of the P$-$O bonds. Among the Al-related impurities, we find that Al$_{\rm Fe}^{+}$ has the lowest formation energy at $\mu_{e}^{\rm int}$ for all thermodynamically allowed Li, Fe, P, O$_{2}$ chemical potentials. The formation energy of the lowest energy configuration of Al$_{\rm Li}$ is found to be higher than that of Al$_{\rm Fe}$ by 0.65$-$0.89 eV, as measured at $\mu_{e}^{\rm int}$.

With Al existing as the positively charged impurity Al$_{\rm Fe}^{+}$, more $V_{\rm Li}^{-}$ and Li$_{\rm Fe}^{-}$ are created to maintain charge neutrality, thus the Fermi level is shifted away from $\mu_{e}^{\rm int}$ and toward the CBM. As a result of this shift, the formation energy (concentration) of $p^{+}$ is increased (decreased). We also investigated possible defect complexes between Al$_{\rm Fe}^{+}$ and the negatively charged native point defects, i.e., Al$_{\rm Fe}^{+}$-$V_{\rm Li}^{-}$ and Al$_{\rm Fe}^{+}$-Li$_{\rm Fe}^{-}$. We find that the binding energy of these complexes are, respectively, 0.38 and 0.25 eV with respect to their isolated constituents. For any given set of the atomic chemical potentials, Al$_{\rm Fe}^{+}$-Li$_{\rm Fe}^{-}$ has a slightly higher (0.04$-$0.23 eV) formation energy than Al$_{\rm Fe}^{+}$-$V_{\rm Li}^{-}$. The distance between the two defects in Al$_{\rm Fe}^{+}$-Li$_{\rm Fe}^{-}$ is 3.28 {\AA}, whereas it is 3.96 {\AA} in Al$_{\rm Fe}^{+}$-Li$_{\rm Fe}^{-}$. Our results suggest that, although Al$_{\rm Fe}^{+}$ enhances both $V_{\rm Li}^{-}$ and Li$_{\rm Fe}^{-}$, the concentration enhancement is larger for the vacancies than for the antisites.

Experimentally, Amin et al.~\cite{B801234B, B801795F} reported that Al-doped LiFePO$_{4}$ has a higher ionic conductivity but lower electronic conductivity compared to undoped LiFePO$_{4}$. This is exactly consistent with our results that Al$_{\rm Fe}^{+}$ enhances $V_{\rm Li}^{-}$ and reduces $p^{+}$. Regarding the lattice site preference, although Amin et al. suggested that Al occurs on the Fe site, which is in agreement with our results, other research groups reported that the dopant is predominantly on the Li site.\cite{Chung:2002p246, Wagemaker2008, meethong2009} Further experimental studies, therefore, should be carried out to clarify this situation.

\subsection{Tetravalent impurities}

\begin{figure}
\begin{center}
\includegraphics[width=3.4in]{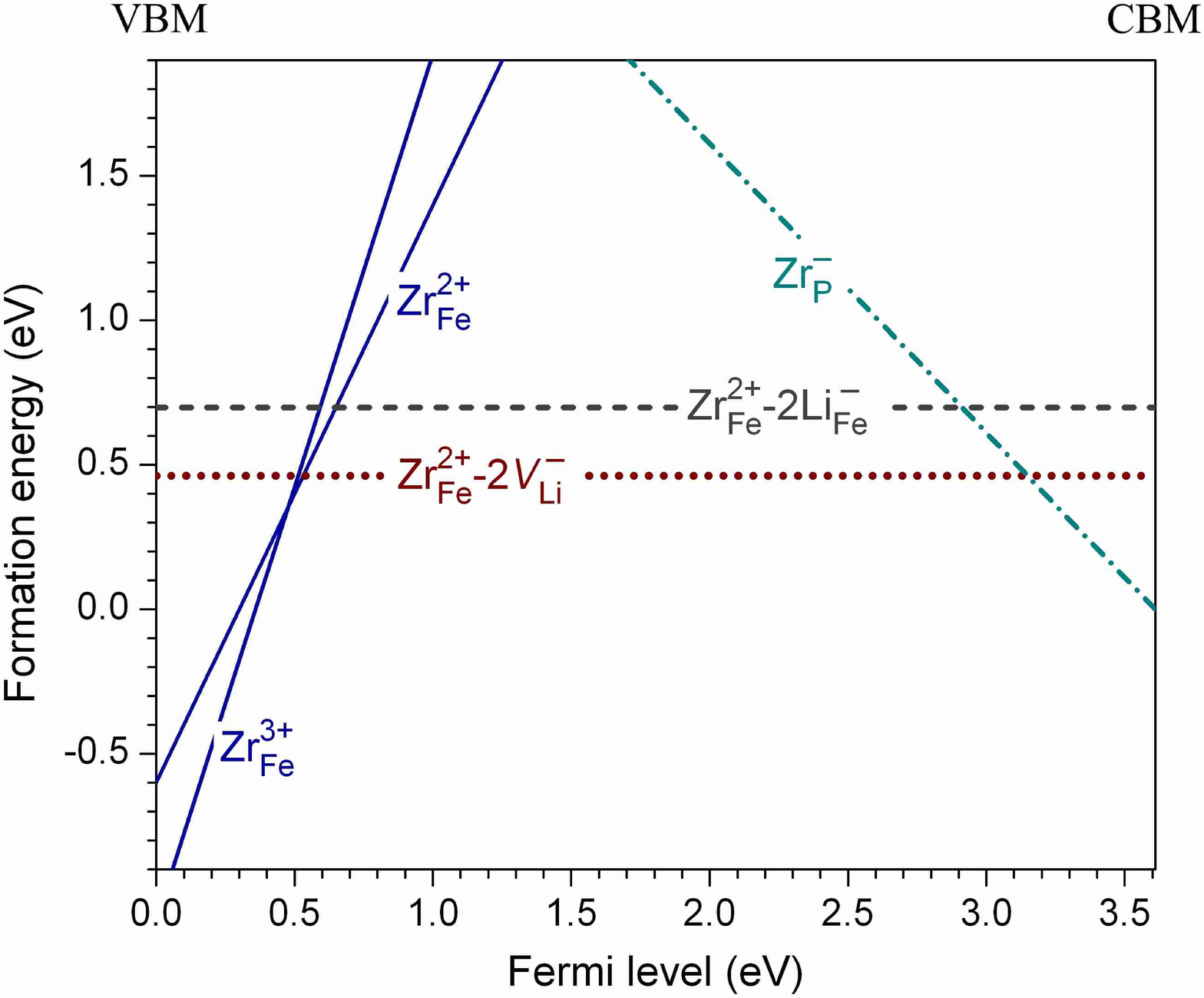}
\end{center}
\vspace{-0.2in}
\caption{Formation energies of Zr-related impurities at $\mu_{{\rm O}_{2}}$=$-$3.03 eV, plotted as a function of Fermi level with respect to the VBM.}\label{fig;Zr}
\end{figure}

Figure \ref{fig;Zr} shows the formation energies of Zr-related substitutional impurities at the Fe and P sites, evaluated at $\mu_{{\rm O}_{2}}$=$-$3.03 eV. Potentially relevant defect configurations are Zr$_{\rm Fe}^{2+}$, Zr$_{\rm Fe}^{3+}$ (a complex of Zr$_{\rm Fe}^{2+}$ and $p^{+}$), and Zr$_{\rm P}^{-}$. In Zr$_{\rm P}^{-}$, Zr replaces P and forms a ZrO$_{4}$ unit with the average Zr$-$O bond length of 1.98 {\AA}, compared to 1.56 {\AA} of the P$-$O bonds. We find that Zr$_{\rm Fe}^{2+}$ has the lowest formation energy at $\mu_{e}^{\rm int}$ for all possible Li, Fe, P, O$_{2}$ chemical potentials, suggesting that Zr has a donor-like effect, i.e., shifting the Fermi level toward CBM and hence enhancing the negatively charged native defects. The formation energy of the lowest energy configuration of Zr$_{\rm Li}$ at $\mu_{e}^{\rm int}$ is found to be higher than that of Zr$_{\rm Fe}$ by 1.17$-$1.41 eV.

We also considered complexes between the elementary Zr$_{\rm Fe}^{2+}$ and the negatively charged native point defects. Zr$_{\rm Fe}^{2+}$-2$V_{\rm Li}^{-}$ is found to have a binding energy of 0.78 eV with respect to its isolated constituents; the distances between Zr$_{\rm Fe}^{2+}$ and the vacancies are 3.26 and 3.62 {\AA}. Zr$_{\rm Fe}^{2+}$-2Li$_{\rm Fe}^{-}$ has a binding energy of 0.65 eV; the distances between Zr$_{\rm Fe}^{2+}$ and the antisites are 4.05 and 4.13 {\AA}. Under Li-deficient environments, the formation energy of Zr$_{\rm Fe}^{2+}$-2$V_{\rm Li}^{-}$ is found to be lower than that of Zr$_{\rm Fe}^{2+}$-2Li$_{\rm Fe}^{-}$ by about 0.20$-$0.34 eV as seen in Fig.~\ref{fig;Zr}; whereas under Li-excess environments, the energy of Zr$_{\rm Fe}^{2+}$-2$V_{\rm Li}^{-}$ is comparable to that of Zr$_{\rm Fe}^{2+}$-2Li$_{\rm Fe}^{-}$. Our results therefore suggest that Zr$_{\rm Fe}^{2+}$ is more likely to be associated with $V_{\rm Li}^{-}$ than with Li$_{\rm Fe}^{-}$, thus enhancing the vacancies and the ionic conductivity.

The donor-like effect of Zr in LiFePO$_{4}$ could have contributed to the enhancement in the total conductivity as observed in experiment.\cite{Chung:2002p246} However, in order to clearly see the effect, one should carefully look at the ionic and electronic conductivities in undoped and doped samples (preferably without carbon coating). Note that, an enhancement in the intrinsic ionic conductivity usually results in a reduction in the intrinsic electronic conductivity, as discussed in the Introduction. Regarding the lattice site preference, several experimental works suggested that Zr occupies the Li site,\cite{Chung:2002p246, Wagemaker2008, meethong2009} which is in contrast to our results showing that the Fe site is energetically more favorable under thermodynamic equilibrium conditions.

\begin{figure}
\begin{center}
\includegraphics[width=3.4in]{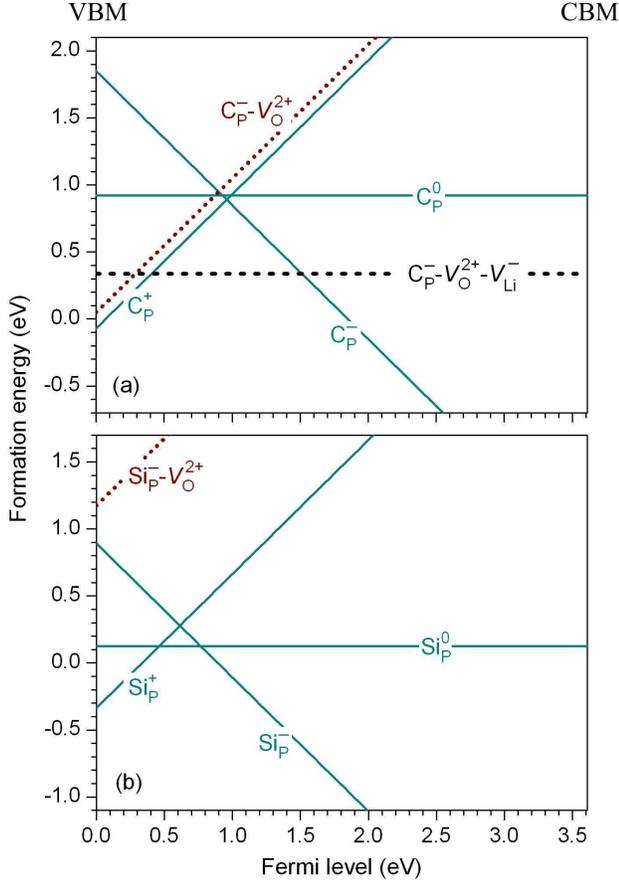}
\end{center}
\vspace{-0.2in}
\caption{Formation energies of (a) C- and (b) Si-related impurities at $\mu_{{\rm O}_{2}}$=$-$3.03 eV, plotted as a function of Fermi level with respect to the VBM.}\label{fig;C-Si}
\end{figure}

Figures \ref{fig;C-Si}(a) and \ref{fig;C-Si}(b) show the formation energies of C- and Si-related substitutional impurities at the P site, evaluated at $\mu_{{\rm O}_{2}}$=$-$3.03 eV. C$_{\rm P}^{-}$ is an elementary defect, whereas C$_{\rm P}^{0}$ is a complex of C$_{\rm P}^{-}$ and $p^{+}$, C$_{\rm P}^{+}$ a complex of C$_{\rm P}^{-}$ and two $p^{+}$. We also considered C$_{\rm P}^{-}$-$V_{\rm O}^{2+}$, a positively charged complex consisting of C$_{\rm P}^{-}$ and an oxygen vacancy ($V_{\rm O}^{2+}$). This complex has a binding energy of 4.44 eV with respect to its isolated constituents. It can also be regarded as replacing a (PO$_{4}$)$^{3-}$ tetrahedron with a trigonal planar (CO$_{3}$)$^{2-}$. In the (CO$_{3}$)$^{2-}$ unit, the average C$-$O bond length is 1.31 {\AA}, compared to 1.56 {\AA} of the P$-$O bonds. C thus exhibits a +4 valence at the P site. At the Li and Fe sites, we however find that C exhibits a +2 valence. As a result, C$_{\rm Li}^{+}$ and C$_{\rm Fe}^{0}$ are the elementary defects, whereas their higher charge states are defect complexes consisting of C$_{\rm Li}^{+}$ (or C$_{\rm Fe}^{0}$) and $p^{+}$.

For $-$3.25 eV$\leq$$\mu_{{\rm O}_{2}}$$\leq$$-$3.03 eV, C$_{\rm P}^{+}$ has the lowest formation energy at $\mu_{e}^{\rm int}$. Given its positive effective charge, this defect would exhibit an donor-like effect, enhancing the concentration of negatively charged native defects. Moreover, for each C$_{\rm P}^{+}$ created, there are two $p^{+}$; thus, with C$_{\rm P}^{+}$ being the lowest energy configuration, C can enhance both $p^{+}$ and $V_{\rm Li}^{-}$ (see more below). For $-$8.21 eV$\leq$$\mu_{{\rm O}_{2}}$$<$$-$3.25 eV, the formation energy of C$_{\rm P}^{-}$-$V_{\rm O}^{2+}$ is lower than that of C$_{\rm P}^{+}$, and the complex has the lowest energy at $\mu_{e}^{\rm int}$. Since C$_{\rm P}^{-}$-$V_{\rm O}^{2+}$ exhibits a donor-like effect, it can enhance $V_{\rm Li}^{-}$. In fact, we have investigated a complex of C$_{\rm P}^{-}$-$V_{\rm O}^{2+}$ and $V_{\rm Li}^{-}$, hereafter denoted as C$_{\rm P}^{-}$-$V_{\rm O}^{2+}$-$V_{\rm Li}^{-}$, and find that it has a binding energy of 0.63 eV with respect to its isolated constituents. The formation energy of this complex [see Fig.~\ref{fig;C-Si}(a)] is lower than that of C$_{\rm P}^{+}$-$V_{\rm Li}^{-}$ (not shown in the figure), a neutral complex of C$_{\rm P}^{+}$ and $V_{\rm Li}^{-}$, by 0.25$-$3.79 eV. Our results suggest that, in the presence of $V_{\rm Li}^{-}$, C$_{\rm P}^{-}$-$V_{\rm O}^{2+}$ is energetically more favorable than C$_{\rm P}^{+}$ for all possible Li, Fe, P, and O$_{2}$ chemical potentials. In other words, if incorporated into the bulk of LiFePO$_{4}$, C is likely to exist as C$_{\rm P}^{-}$-$V_{\rm O}^{2+}$-$V_{\rm Li}^{-}$, and the presence of C thus enhances $V_{\rm Li}^{-}$ only. The formation energy of C at the Li and Fe sites is much higher than at the P site, with the energy difference at $\mu_{e}^{\rm int}$ between C$_{\rm P}$ and C$_{\rm Fe}$ is 4.80$-$6.24 eV. Note that, here we assume C can go into the bulk with a concentration higher than that of the native defects. The penetration of C, however, may be impeded by reactions at the surface.\cite{C0JM04190D} Since C is present in precursors used in the synthesis of LiFePO$_{4}$ and also because LiFePO$_{4}$ is usually coated with C, it is worth carrying out further studies to see if there is C in the interior of the material.

Regarding Si-related substitutional impurities at the P site, Si$_{\rm P}^{-}$ is found to be an elementary defect, whereas Si$_{\rm P}^{0}$ is a complex of Si$_{\rm P}^{-}$ and $p^{+}$, and Si$_{\rm P}^{+}$ a complex of Si$_{\rm P}^{-}$ and two $p^{+}$. Si$_{\rm P}^{-}$-$V_{\rm O}^{2+}$ has a binding energy of 2.35 eV with respect to Si$_{\rm P}^{-}$ and $V_{\rm O}^{2+}$. In this complex, Si forms a SiO$_{4}$ unit that shares one O atom with the neighboring PO$_{4}$ unit, i.e., SiO$_{3}$-O-PO$_{3}$ with the Si$-$P distance being 2.90 {\AA}, compared to 3.71 {\AA} of the P$-$P distance. Energetically, we find that Si$_{\rm P}^{+}$ has a lower formation energy than Si$_{\rm P}^{-}$-$V_{\rm O}^{2+}$ except for $-$8.21 eV$\leq$$\mu_{{\rm O}_{2}}$$\leq$$-$7.59 eV. Si$_{\rm P}^{0}$ is found to have the lowest formation energy at $\mu_{e}^{\rm int}$ for $-$4.59 eV$<$$\mu_{{\rm O}_{2}}$$\leq$$-$3.03 eV, as seen in Fig.~\ref{fig;C-Si}(b) for $\mu_{{\rm O}_{2}}$=$-$3.03 eV. Although Si$_{\rm P}^{0}$ is neutral and thus not effective in shifting the Fermi level, each Si$_{\rm P}^{0}$ contains one $p^{+}$. The formation of Si$_{\rm P}^{0}$, therefore, may help introduce more $p^{+}$ into the material. For $-$8.21 eV$\leq$$\mu_{{\rm O}_{2}}$$\leq$$-$4.59 eV, Si$_{\rm P}^{-}$ is found to have the lowest formation energy. In this case, Si exhibits an acceptor-like effect and is expected to enhance the concentration of $p^{+}$. For any given set of Li, Fe, P, and O$_{2}$ chemical potentials, neutral complexes such as those consisting of $V_{\rm Li}^{-}$ and Si$_{\rm P}^{+}$ or Si$_{\rm P}^{-}$-$V_{\rm O}^{2+}$ all have higher formation energies than Si$_{\rm P}^{0}$. Our results thus suggest that Si incorporated at the P site enhances small hole polarons. At the Li and Fe sites, Si exhibits a donor-like effect since Si is supervalent to Li and Fe; the elementary, lowest energy configurations at $\mu_{e}^{\rm int}$ are Si$_{\rm Li}^{3+}$ and Si$_{\rm Fe}^{2+}$, where the formation energy of Si at the Li site is higher than that at the Fe site by 0.43$-$0.65 eV, depending on the specific set of the chemical potentials. The formation energy difference at $\mu_{e}^{\rm int}$ between Si$_{\rm P}$ and Si$_{\rm Fe}$ is 0.33$-$3.56 eV

Amin et al.~\cite{Amin_Si} reported that doping LiFePO$_{4}$ with Si results in an increase of the ionic conductivity and a decrease of the electronic conductivity, suggesting a donor-like effect. As discussed above, this effect should be associated with Si$_{\rm Fe}^{2+}$. Our results, however, suggest that Si$_{\rm Fe}^{2+}$ is not the lowest defect configuration under thermodynamic equilibrium conditions. Although the cause of this discrepancy between experiment and our results on the lattice site preference of Si is still not clear, we suspect that doping at the P site may be kinetically hindered, and thus a thermodynamic equilibrium approach is not applicable.\cite{walle:3851} This may be the case for all dopants that involve P-site doping in LiFePO$_{4}$. It should be noted that LiFePO$_{4}$ can be regarded as consisting of Li$^{+}$, Fe$^{2+}$, and (PO$_{4}$)$^{3-}$ with ionic bonding between the units. Within the (PO$_{4}$)$^{3-}$ unit, however, the P$-$O bonds are highly covalent.\cite{hoang2011} If doping occurs via diffusion, P-site doping would involve diffusion of P$^{5+}$ and the dopant which is expected to be difficult because it necessarily involves breaking/forming of chemical bonds between O and P (and the dopant). Thus P-site doping might also be affected by the P- and dopant-containing precursors used in the synthesis.

\subsection{Pentavalent impurities}

\begin{figure}
\begin{center}
\includegraphics[width=3.4in]{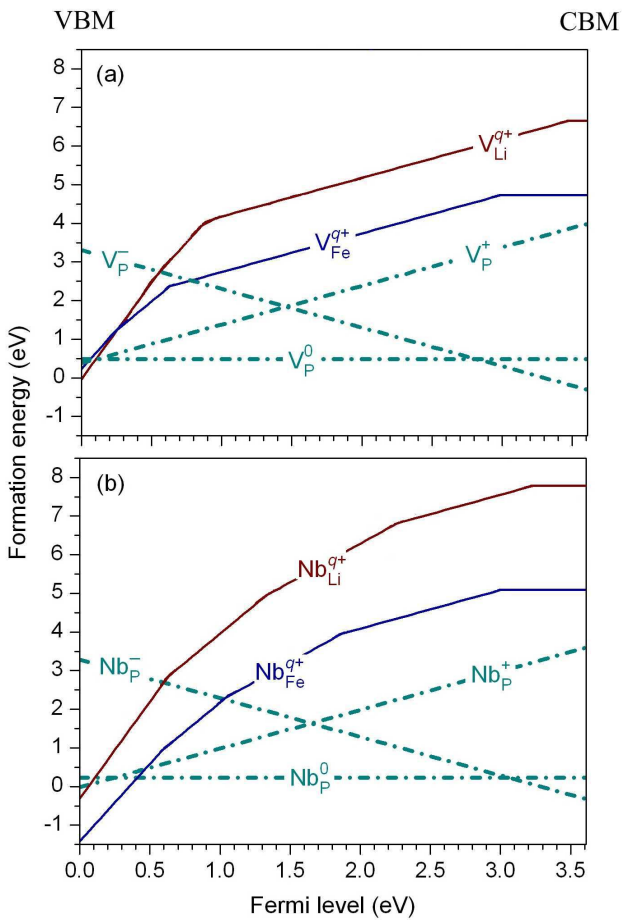}
\end{center}
\vspace{-0.2in}
\caption{Formation energies of (a) V- and (b) Nb-related impurities at $\mu_{{\rm O}_{2}}$=$-$3.03 eV, plotted as a function of Fermi level with respect to the VBM. For substitutional impurities at the Li and Fe sites (M$_{\rm Li}^{q+}$ and M$_{\rm Fe}^{q+}$, where M=V and Nb, and $q$$\geq$0), only the lowest formation-energy segments are presented.}\label{fig;V-Nb}
\end{figure}

Figures \ref{fig;V-Nb}(a) and \ref{fig;V-Nb}(b) show the formation energies of V- and Nb-related substitutional impurities at the Li, Fe, and P sites, evaluated at $\mu_{{\rm O}_{2}}$=$-$3.03 eV. We find that V$_{\rm P}^{0}$ is an elementary point defect, suggesting that V (vanadium, not to be confused with a vacancy, $V$) at the P site has a +5 valence. V$_{\rm P}^{+}$ is a complex of V$_{\rm P}^{0}$ and $p^{+}$, whereas V$_{\rm P}^{-}$ is a complex of V$_{\rm P}^{0}$ and a small electron polaron ($p^-$).\cite{hoang2011} At the Li and Fe sites, however, V exhibits a +3 valence, with V$_{\rm Li}^{2+}$ and V$_{\rm Fe}^{+}$ being the elementary defects. Attempts to create higher charge states result in complexes consisting of these defects and $p^{+}$, whereas lower charge states are complexes containing $p^{-}$. We have also investigated neutral complexes such as V$_{\rm Fe}^{+}$-$V_{\rm Li}^{-}$, a complex of V$_{\rm Fe}^{+}$ and a Li vacancy, and 2V$_{\rm Fe}^{+}$-$V_{\rm Fe}^{2-}$, a complex of two V$_{\rm Fe}^{+}$ and an Fe vacancy. Energetically, we find that V$_{\rm P}^{0}$ has the lowest formation energy at $\mu_{e}^{\rm int}$ for all possible Li, Fe, P, and O$_{2}$ chemical potentials. With this neutral configuration, V would be ineffective in shifting the Fermi level. 

V has positive effective charges at the Li and Fe sites and thus exhibits a donor-like effect. The formation energy at $\mu_{e}^{\rm int}$ of V$_{\rm Fe}$ in its lowest energy configuration is higher than that of V$_{\rm P}^{0}$ by 0.50$-$2.14 eV for almost all possible chemical potentials, and by 0.10$-$0.20 eV for a small region of the chemical potentials associated with $\mu_{{\rm O}_{2}}$ around $-$4.59 eV. On the other hand, the formation energy of V$_{\rm Fe}$ is lower than that of V$_{\rm Li}$ by 0.76$-$0.97 eV for almost all possible chemical potentials, except for a small region associated with $\mu_{{\rm O}_{2}}$$\leq$$-$4.59 eV where the two defects have comparable formation energies. Neutral complexes such as V$_{\rm Fe}^{+}$-$V_{\rm Li}^{-}$ and 2V$_{\rm Fe}^{+}$-$V_{\rm Fe}^{2-}$ have formation energies higher than V$_{\rm P}^{0}$ by 0.36$-$2.32 and 4.51$-$9.09 eV, respectively. Again, these results suggest that V is energetically more favorable at the P site under thermodynamic equilibrium conditions.

Experimentally, Hong et al.~\cite{hong:A33} reported that their attempt to incorporate V into LiFePO$_{4}$ at the Fe site resulted in having V at the P site instead. However, Omenya et al.~\cite{Omenya2011} later reported that the substitution at the P site could not be reproduced, but at least 10 mol\% of the Fe sites were occupied by V$^{3+}$. Zhang et al.~\cite{Zhang2011} also reported a valence between +3 and +4 for V in LiFePO$_{4}$. Note that, according our results dicussed above, the +3 valence of V is associated with substitution at the Fe site that exhibits a donor-like effect. Thus, for V-doped LiFePO$_{4}$ samples reported in Refs.~\cite{Omenya2011, Zhang2011}, V may indeed have been incorporated at the Fe site, although this site is found to be energetically less favorable in our calculations. The discrepancy between experiment and our results on the lattice site preference of V could be that P-site doping is thermodynamically favorable but kinetically hindered, as discussed earlier.

Regarding Nb, we find that Nb$_{\rm P}^{0}$ is an elementary point defect, and the dopant thus has a +5 valence at the P site. Nb$_{\rm P}^{+}$ is a complex of Nb$_{\rm P}^{0}$ and $p^{+}$, whereas Nb$_{\rm P}^{-}$ is a complex of Nb$_{\rm P}^{0}$ and $p^-$. Nb also exhibits the +5 valence at the Fe site, with Nb$_{\rm Fe}^{3+}$ being the elementary defect. At the Li site, we however find that Nb has a +4 valence and the elementary configuration is Nb$_{\rm Li}^{3+}$. Higher charge states such as Nb$_{\rm Li}^{4+}$ is a complex of Nb$_{\rm Li}^{3+}$ and $p^{+}$. Energetically, Nb$_{\rm P}^{0}$ has the lowest formation energy at $\mu_{e}^{\rm int}$ for almost all possible chemical potentials, except for a very small region associated with $-$4.59 eV$\leq$$\mu_{{\rm O}_{2}}$$\leq$$-$3.89 eV of the ($\mu_{\rm Li}$, $\mu_{\rm Fe}$, $\mu_{{\rm O}_{2}}$) polyhedron that defines the stable range of Li, Fe, P, and O$_{2}$ chemical potentials (Ref.~\cite{hoang2011}) where Nb$_{\rm Fe}^{3+}$ has the lowest formation energy. For example, in the $\mu_{{\rm O}_{2}}$=$-$4.59 eV plane, the small region near Points B, C, and D in the chemical-potential diagram as depicted in Fig.~1 of Ref.~\cite{hoang2011} is where Nb$_{\rm Fe}^{3+}$ is most favorable energetically. Nb$_{\rm P}^{0}$ cannot enhance the concentration of either $V_{\rm Li}^{-}$ or $p^{+}$, but Nb$_{\rm Fe}^{3+}$ is donor-like and thus can enhance $V_{\rm Li}^{-}$. We also find that Nb$_{\rm Fe}^{3+}$-3$V_{\rm Li}^{-}$, a complex of Nb$_{\rm Fe}^{3+}$ and three Li vacancies, has a lower formation energy than Nb$_{\rm P}^{0}$ in that small region of the ($\mu_{\rm Li}$, $\mu_{\rm Fe}$, $\mu_{{\rm O}_{2}}$) polyhedron. Our results thus suggest that Nb can be incorporated at the Fe or P site, depending on the synthesis conditions. The reported enhancement in the total conductivity of Nb-doped LiFePO$_{4}$ (Ref.~\cite{Chung:2002p246}) could have been partly due to the enhancement of $V_{\rm Li}^{-}$ caused by Nb$_{\rm Fe}^{3+}$. In addition, the incorporation of Nb at the P site may be kinetically hindered as discussed above in the case of Si and V.

Overall, we find that extrinsic impurities in LiFePO$_{4}$ each have one stable charge state (i.e., elementary defect configuration) at a given lattice site. Any attempt to create higher (lower) charge states will result in complexes consisting of the elementary configuration and small hole (electron) polarons, which is similar to what has been observed for native defects in LiFePO$_{4}$ as reported in Ref.~\cite{hoang2011}. Our studies help identify specific impurities that are likely to be effective (or ineffective) for enhancing the concentration of $p^{+}$ or $V_{\rm Li}^{-}$. The results also suggest further experiments to investigate the lattice site preference (and its dependence on the synthesis conditions) and changes in the ionic and electronic conductivities. For practical applications, since doping alone cannot enhance both $p^{+}$ and $V_{\rm Li}^{-}$ as needed for high ionic and electronic conductivities, other methods such as carbon coating (Refs.~\cite{huang:A170,Ellis2007}) and/or thermal treatment (Ref.~\cite{Amin20081831}) may still be needed. As discussed by Julien et al.~\cite{C0JM04190D}, carbon coating can help repair structural damage at the surface and ensure the electric contact between LiFePO$_{4}$ particles. Also, with whatever method is employed to enhance the conductivity, one should first be able to control the experimental conditions during synthesis to reduce Fe$_{\rm Li}^{+}$ and enhance $p^{+}$ and $V_{\rm Li}^{-}$.\cite{hoang2011}

Note that computational studies of impurities in LiFePO$_{4}$ have also been carried out by Islam et al.~\cite{Islam:2005p168, Fisher:2008p80}. Based on calculations using interatomic potentials, they reported that, among other impurities, Na is energetically more favorable at the Li site, whereas Mg, Al, Zr, and Nb are more favorable at the Fe site. These results are generally in qualitative agreement with our results for these substitutional impurities at the Li and Fe sites. However, they did not report results for the impurities at the P site, and also did not explore all possible Li, Fe, P, and O$_{2}$ chemical potentials. As discussed earlier, the lattice site preference can be sensitive to the chemical potentials (e.g., in the case of K and Cu), and the Li and Fe sites may not be energetically most favorable under thermodynamic equilibrium.

\section{\label{sec;summary}Summary}

We have carried out first-principles studies of the effects of various extrinsic impurities on the ionic and electronic conduction in LiFePO$_{4}$. We find that the formation energy and lattice site preference of the impurities depend on Li, Fe, P, and O$_{2}$ chemical potentials which represent the synthesis conditions. For all thermodynamically allowed atomic chemical potentials, Na, Cu, Ag, Mg, and Zn are likely to exist as neutral defects in LiFePO$_{4}$ and thus do not enhance the concentration of either small hole polarons ($p^{+}$) or lithium vacancies ($V_{\rm Li}^{-}$), i.e. electronic or ionic conduction. K may be incorporated at the Fe site under thermodynamic equilibrium for a small range of the chemical potentials, where it exhibits an acceptor-like effect and thus can enhance the concentration of $p^{+}$. Al and Zr, on the other hand, exhibit a donor-like effect and are thus effective in enhancing the concentration of $V_{\rm Li}^{-}$. C can also exhibit a donor-like effect if incorporated into the material. We find that Si is energetically more favorable at the P site where it exhibits an acceptor-like effect, which is in contrast to experiment suggesting substitution at the Fe site that exhibits a donor-like effect. V is found to be more favorable at the P site and thus electrically inactive, in contrast to experiment suggesting substitution at the Fe site. Nb is also found to be electrically inactive at the P site, except for a small range of the atomic chemical potentials where it is energetically more favorable at the Fe site and exhibits a donor-like effect. We suggest that doping LiFePO$_{4}$ with Si, V, or Nb at the P site is thermodynamically favorable but kinetically hindered, and that Si, V, and Nb may have been incorporated at the Fe site where they exhibit a donor-like effect as reported in experiment. Indeed, the valence reported for V in V-doped LiFePO$_{4}$ by experiment is consistent with substitution at the Fe site in our calculations. Our studies, therefore, can serve as guidelines for experiment on which lattice site the impurities are most likely to be incorporated under thermodynamic equilibrium conditions, and how the impurities may affect the ionic and electronic conduction. To achieve a high electrical conductivity, however, one should combine the incorporation of electrically active impurities with other methods such as defect-controlled synthesis, carbon coating, and/or thermal treatment.

\begin{acknowledgments}

We acknowledge helpful discussions with S.~C.~Erwin, C.~S.~Hellberg, and J.~Allen, and the use of computing facilities at the DoD HPC Centers. K.~H.~was supported by the U.S. Naval Research Laboratory through Grant No.~NRL-N00173-08-G001, and M.~D.~J.~by the Office of Naval Research.

\end{acknowledgments}



\begin{thebibliography}{49}
\expandafter\ifx\csname natexlab\endcsname\relax\def\natexlab#1{#1}\fi
\providecommand{\bibinfo}[2]{#2}
\ifx\xfnm\relax \def\xfnm[#1]{\unskip,\space#1}\fi
\bibitem[{Padhi et~al.(1997)Padhi, Nanjundaswamy, and Goodenough}]{padhi:1188}
\bibinfo{author}{A.~K. Padhi}, \bibinfo{author}{K.~S. Nanjundaswamy},
  \bibinfo{author}{J.~B. Goodenough}, \bibinfo{journal}{J. Electrochem. Soc.}
  \bibinfo{volume}{144} (\bibinfo{year}{1997}) \bibinfo{pages}{1188--1194}.
\bibitem[{Ellis et~al.(2010)Ellis, Lee, and Nazar}]{Ellis2010}
\bibinfo{author}{B.~L. Ellis}, \bibinfo{author}{K.~T. Lee},
  \bibinfo{author}{L.~F. Nazar}, \bibinfo{journal}{Chem. Mater.}
  \bibinfo{volume}{22} (\bibinfo{year}{2010}) \bibinfo{pages}{691--714}.
\bibitem[{Manthiram(2011)}]{Manthiram2011}
\bibinfo{author}{A.~Manthiram}, \bibinfo{journal}{J. Phys. Chem. Lett.}
  \bibinfo{volume}{2} (\bibinfo{year}{2011}) \bibinfo{pages}{176--184}.
\bibitem[{Chung et~al.(2002)Chung, Bloking, and Chiang}]{Chung:2002p246}
\bibinfo{author}{S.~Chung}, \bibinfo{author}{J.~Bloking},
  \bibinfo{author}{Y.~Chiang}, \bibinfo{journal}{Nat. Mater.}
  \bibinfo{volume}{1} (\bibinfo{year}{2002}) \bibinfo{pages}{123--128}.
\bibitem[{Ravet et~al.(2003)Ravet, Abouimrane, and Armand}]{Ravet2003}
\bibinfo{author}{N.~Ravet}, \bibinfo{author}{A.~Abouimrane},
  \bibinfo{author}{M.~Armand}, \bibinfo{journal}{{Nat. Mater.}}
  \bibinfo{volume}{{2}} (\bibinfo{year}{{2003}}) \bibinfo{pages}{{702}}.
\bibitem[{Wagemaker et~al.(2008)Wagemaker, Ellis, L\"{u}tzenkirchen-Hecht,
  Mulder, and Nazar}]{Wagemaker2008}
\bibinfo{author}{M.~Wagemaker}, \bibinfo{author}{B.~L. Ellis},
  \bibinfo{author}{D.~L\"{u}tzenkirchen-Hecht}, \bibinfo{author}{F.~M. Mulder},
  \bibinfo{author}{L.~F. Nazar}, \bibinfo{journal}{Chem. Mater.}
  \bibinfo{volume}{20} (\bibinfo{year}{2008}) \bibinfo{pages}{6313--6315}.
\bibitem[{Hoang and Johannes(2011)}]{hoang2011}
\bibinfo{author}{K.~Hoang}, \bibinfo{author}{M.~Johannes},
  \bibinfo{journal}{Chem. Mater.} \bibinfo{volume}{23} (\bibinfo{year}{2011})
  \bibinfo{pages}{3003--3013}.
\bibitem[{Axmann et~al.(2009)Axmann, Stinner, Wohlfahrt-Mehrens, Mauger,
  Gendron, and Julien}]{Axmann:2009p137}
\bibinfo{author}{P.~Axmann}, \bibinfo{author}{C.~Stinner},
  \bibinfo{author}{M.~Wohlfahrt-Mehrens}, \bibinfo{author}{A.~Mauger},
  \bibinfo{author}{F.~Gendron}, \bibinfo{author}{C.~M. Julien},
  \bibinfo{journal}{Chem. Mater.} \bibinfo{volume}{21} (\bibinfo{year}{2009})
  \bibinfo{pages}{1636--1644}.
\bibitem[{Chung et~al.(2010)Chung, Kim, and Choi}]{Chung2010}
\bibinfo{author}{S.-Y. Chung}, \bibinfo{author}{Y.-M. Kim},
  \bibinfo{author}{S.-Y. Choi}, \bibinfo{journal}{Adv. Funct. Mater.}
  \bibinfo{volume}{20} (\bibinfo{year}{2010}) \bibinfo{pages}{4219--4232}.
\bibitem[{Zhou et~al.(2004)Zhou, Kang, Maxisch, Ceder, and
  Morgan}]{Zhou:2004p101}
\bibinfo{author}{F.~Zhou}, \bibinfo{author}{K.~Kang},
  \bibinfo{author}{T.~Maxisch}, \bibinfo{author}{G.~Ceder},
  \bibinfo{author}{D.~Morgan}, \bibinfo{journal}{Solid State Commun.}
  \bibinfo{volume}{132} (\bibinfo{year}{2004}) \bibinfo{pages}{181--186}.
\bibitem[{Maxisch et~al.(2006)Maxisch, Zhou, and Ceder}]{Maxisch:2006p103}
\bibinfo{author}{T.~Maxisch}, \bibinfo{author}{F.~Zhou},
  \bibinfo{author}{G.~Ceder}, \bibinfo{journal}{Phys. Rev. B}
  \bibinfo{volume}{73} (\bibinfo{year}{2006}) \bibinfo{pages}{104301}.
\bibitem[{Ellis et~al.(2006)Ellis, Perry, Ryan, and Nazar}]{Ellis2006}
\bibinfo{author}{B.~Ellis}, \bibinfo{author}{L.~K. Perry},
  \bibinfo{author}{D.~H. Ryan}, \bibinfo{author}{L.~F. Nazar},
  \bibinfo{journal}{J. Am. Chem. Soc.} \bibinfo{volume}{128}
  (\bibinfo{year}{2006}) \bibinfo{pages}{11416--11422}.
\bibitem[{Zaghib et~al.(2007)Zaghib, Mauger, Goodenough, Gendron, and
  Julien}]{Zaghib2007}
\bibinfo{author}{K.~Zaghib}, \bibinfo{author}{A.~Mauger},
  \bibinfo{author}{J.~B. Goodenough}, \bibinfo{author}{F.~Gendron},
  \bibinfo{author}{C.~M. Julien}, \bibinfo{journal}{Chem. Mater.}
  \bibinfo{volume}{19} (\bibinfo{year}{2007}) \bibinfo{pages}{3740--3747}.
\bibitem[{Peles and {Van de Walle}(2007)}]{peles2007}
\bibinfo{author}{A.~Peles}, \bibinfo{author}{C.~G. {Van de Walle}},
  \bibinfo{journal}{Phys. Rev. B} \bibinfo{volume}{76} (\bibinfo{year}{2007})
  \bibinfo{pages}{214101}.
\bibitem[{Hoang and {Van de Walle}(2009)}]{hoang2009}
\bibinfo{author}{K.~Hoang}, \bibinfo{author}{C.~G. {Van de Walle}},
  \bibinfo{journal}{Phys. Rev. B} \bibinfo{volume}{80} (\bibinfo{year}{2009})
  \bibinfo{pages}{214109}.
\bibitem[{Wilson-Short et~al.(2009)Wilson-Short, Janotti, Hoang, Peles, and
  {Van de Walle}}]{wilson-short}
\bibinfo{author}{G.~B. Wilson-Short}, \bibinfo{author}{A.~Janotti},
  \bibinfo{author}{K.~Hoang}, \bibinfo{author}{A.~Peles},
  \bibinfo{author}{C.~G. {Van de Walle}}, \bibinfo{journal}{Phys. Rev. B}
  \bibinfo{volume}{80} (\bibinfo{year}{2009}) \bibinfo{pages}{224102}.
\bibitem[{Hoang et~al.(2011)Hoang, Janotti, and Van~de
  Walle}]{hoang_amide_angew}
\bibinfo{author}{K.~Hoang}, \bibinfo{author}{A.~Janotti},
  \bibinfo{author}{C.~G. Van~de Walle}, \bibinfo{journal}{Angew. Chem. Int.
  Ed.} \bibinfo{volume}{50} (\bibinfo{year}{2011})
  \bibinfo{pages}{10170--10173}.
\bibitem[{{Van de Walle} and Neugebauer(2004)}]{walle:3851}
\bibinfo{author}{C.~G. {Van de Walle}}, \bibinfo{author}{J.~Neugebauer},
  \bibinfo{journal}{J. Appl. Phys.} \bibinfo{volume}{95} (\bibinfo{year}{2004})
  \bibinfo{pages}{3851--3879}.
\bibitem[{Anisimov et~al.(1991)Anisimov, Zaanen, and Andersen}]{anisimov1991}
\bibinfo{author}{V.~I. Anisimov}, \bibinfo{author}{J.~Zaanen},
  \bibinfo{author}{O.~K. Andersen}, \bibinfo{journal}{Phys. Rev. B}
  \bibinfo{volume}{44} (\bibinfo{year}{1991}) \bibinfo{pages}{943--954}.
\bibitem[{Anisimov et~al.(1993)Anisimov, Solovyev, Korotin,
  Czy\ifmmode~\dot{z}\else \.{z}\fi{}yk, and Sawatzky}]{anisimov1993}
\bibinfo{author}{V.~I. Anisimov}, \bibinfo{author}{I.~V. Solovyev},
  \bibinfo{author}{M.~A. Korotin}, \bibinfo{author}{M.~T.
  Czy\ifmmode~\dot{z}\else \.{z}\fi{}yk}, \bibinfo{author}{G.~A. Sawatzky},
  \bibinfo{journal}{Phys. Rev. B} \bibinfo{volume}{48} (\bibinfo{year}{1993})
  \bibinfo{pages}{16929--16934}.
\bibitem[{Liechtenstein et~al.(1995)Liechtenstein, Anisimov, and
  Zaanen}]{liechtenstein1995}
\bibinfo{author}{A.~I. Liechtenstein}, \bibinfo{author}{V.~I. Anisimov},
  \bibinfo{author}{J.~Zaanen}, \bibinfo{journal}{Phys. Rev. B}
  \bibinfo{volume}{52} (\bibinfo{year}{1995}) \bibinfo{pages}{R5467--R5470}.
\bibitem[{Perdew et~al.(1996)Perdew, Burke, and Ernzerhof}]{GGA}
\bibinfo{author}{J.~P. Perdew}, \bibinfo{author}{K.~Burke},
  \bibinfo{author}{M.~Ernzerhof}, \bibinfo{journal}{Phys. Rev. Lett.}
  \bibinfo{volume}{77} (\bibinfo{year}{1996}) \bibinfo{pages}{3865--3868}.
\bibitem[{Bl\"ochl(1994)}]{PAW1}
\bibinfo{author}{P.~E. Bl\"ochl}, \bibinfo{journal}{Phys. Rev. B}
  \bibinfo{volume}{50} (\bibinfo{year}{1994}) \bibinfo{pages}{17953--17979}.
\bibitem[{Kresse and Joubert(1999)}]{PAW2}
\bibinfo{author}{G.~Kresse}, \bibinfo{author}{D.~Joubert},
  \bibinfo{journal}{Phys. Rev. B} \bibinfo{volume}{59} (\bibinfo{year}{1999})
  \bibinfo{pages}{1758--1775}.
\bibitem[{Kresse and Hafner(1993)}]{VASP1}
\bibinfo{author}{G.~Kresse}, \bibinfo{author}{J.~Hafner},
  \bibinfo{journal}{Phys. Rev. B} \bibinfo{volume}{47} (\bibinfo{year}{1993})
  \bibinfo{pages}{558--561}.
\bibitem[{Kresse and Furthm\"uller(1996{\natexlab{a}})}]{VASP2}
\bibinfo{author}{G.~Kresse}, \bibinfo{author}{J.~Furthm\"uller},
  \bibinfo{journal}{Phys. Rev. B} \bibinfo{volume}{54}
  (\bibinfo{year}{1996}{\natexlab{a}}) \bibinfo{pages}{11169--11186}.
\bibitem[{Kresse and Furthm\"uller(1996{\natexlab{b}})}]{VASP3}
\bibinfo{author}{G.~Kresse}, \bibinfo{author}{J.~Furthm\"uller},
  \bibinfo{journal}{Comput. Mat. Sci.} \bibinfo{volume}{6}
  (\bibinfo{year}{1996}{\natexlab{b}}) \bibinfo{pages}{15--50}.
\bibitem[{Zhou et~al.(2004)Zhou, Cococcioni, Marianetti, Morgan, and
  Ceder}]{Zhou:2004p104}
\bibinfo{author}{F.~Zhou}, \bibinfo{author}{M.~Cococcioni},
  \bibinfo{author}{C.~Marianetti}, \bibinfo{author}{D.~Morgan},
  \bibinfo{author}{G.~Ceder}, \bibinfo{journal}{Phys. Rev. B}
  \bibinfo{volume}{70} (\bibinfo{year}{2004}) \bibinfo{pages}{235121}.
\bibitem[{Ong et~al.(2008)Ong, Wang, Kang, and Ceder}]{ong2008}
\bibinfo{author}{P.~S. Ong}, \bibinfo{author}{L.~Wang},
  \bibinfo{author}{B.~Kang}, \bibinfo{author}{G.~Ceder},
  \bibinfo{journal}{Chem. Mater.} \bibinfo{volume}{20} (\bibinfo{year}{2008})
  \bibinfo{pages}{1798--1807}.
\bibitem[{Reuter and Scheffler(2001)}]{Reuter2001}
\bibinfo{author}{K.~Reuter}, \bibinfo{author}{M.~Scheffler},
  \bibinfo{journal}{Phys. Rev. B} \bibinfo{volume}{65} (\bibinfo{year}{2001})
  \bibinfo{pages}{035406}.
\bibitem[{Yin et~al.(2010)Yin, Huang, Liu, Wang, and Wang}]{Yin20104308}
\bibinfo{author}{X.~Yin}, \bibinfo{author}{K.~Huang}, \bibinfo{author}{S.~Liu},
  \bibinfo{author}{H.~Wang}, \bibinfo{author}{H.~Wang}, \bibinfo{journal}{J.
  Power Sources} \bibinfo{volume}{195} (\bibinfo{year}{2010})
  \bibinfo{pages}{4308 -- 4312}.
\bibitem[{Fang et~al.(2011)Fang, Li, Huang, Liu, Huang, Zhuang, and
  Zhang}]{fang_kdoped}
\bibinfo{author}{X.~Fang}, \bibinfo{author}{J.~Li}, \bibinfo{author}{K.~Huang},
  \bibinfo{author}{S.~Liu}, \bibinfo{author}{C.~Huang},
  \bibinfo{author}{S.~Zhuang}, \bibinfo{author}{J.~Zhang}, \bibinfo{journal}{J.
  Solid State Electrochem.}  (\bibinfo{year}{2011}),
  \bibinfo{note}{doi: 10.1007/s10008-011-1426-4}.
\bibitem[{Roberts et~al.(2008)Roberts, Vitins, and Owen}]{Roberts2008754}
\bibinfo{author}{M.~R. Roberts}, \bibinfo{author}{G.~Vitins},
  \bibinfo{author}{J.~R. Owen}, \bibinfo{journal}{J. Power Sources}
  \bibinfo{volume}{179} (\bibinfo{year}{2008}) \bibinfo{pages}{754 -- 762}.
\bibitem[{Guo et~al.(2005)Guo, Liu, Bewlay, Liu, and Dou}]{Guo2005113}
\bibinfo{author}{Z.~Guo}, \bibinfo{author}{H.~Liu},
  \bibinfo{author}{S.~Bewlay}, \bibinfo{author}{H.~Liu},
  \bibinfo{author}{S.~Dou}, \bibinfo{journal}{Synthetic Metals}
  \bibinfo{volume}{153} (\bibinfo{year}{2005}) \bibinfo{pages}{113--116}.
\bibitem[{Meethong et~al.(2009)Meethong, Kao, Speakman, and
  Chiang}]{meethong2009}
\bibinfo{author}{N.~Meethong}, \bibinfo{author}{Y.-H. Kao},
  \bibinfo{author}{S.~A. Speakman}, \bibinfo{author}{Y.-M. Chiang},
  \bibinfo{journal}{Adv. Funct. Mater.} \bibinfo{volume}{19}
  (\bibinfo{year}{2009}) \bibinfo{pages}{1060--1070}.
\bibitem[{Wang and Hong(2007)}]{wang:A65}
\bibinfo{author}{C.~Wang}, \bibinfo{author}{J.~Hong},
  \bibinfo{journal}{Electrochem. Solid-State Lett.} \bibinfo{volume}{10}
  (\bibinfo{year}{2007}) \bibinfo{pages}{A65--A69}.
\bibitem[{Bilecka et~al.(2011)Bilecka, Hintennach, Rossell, Xie, Novak, and
  Niederberger}]{C0JM03476B}
\bibinfo{author}{I.~Bilecka}, \bibinfo{author}{A.~Hintennach},
  \bibinfo{author}{M.~D. Rossell}, \bibinfo{author}{D.~Xie},
  \bibinfo{author}{P.~Novak}, \bibinfo{author}{M.~Niederberger},
  \bibinfo{journal}{J. Mater. Chem.} \bibinfo{volume}{21}
  (\bibinfo{year}{2011}) \bibinfo{pages}{5881--5890}.
\bibitem[{Amin et~al.(2008{\natexlab{a}})Amin, Lin, and Maier}]{B801234B}
\bibinfo{author}{R.~Amin}, \bibinfo{author}{C.~Lin},
  \bibinfo{author}{J.~Maier}, \bibinfo{journal}{Phys. Chem. Chem. Phys.}
  \bibinfo{volume}{10} (\bibinfo{year}{2008}{\natexlab{a}})
  \bibinfo{pages}{3519--3523}.
\bibitem[{Amin et~al.(2008{\natexlab{b}})Amin, Lin, and Maier}]{B801795F}
\bibinfo{author}{R.~Amin}, \bibinfo{author}{C.~Lin},
  \bibinfo{author}{J.~Maier}, \bibinfo{journal}{Phys. Chem. Chem. Phys.}
  \bibinfo{volume}{10} (\bibinfo{year}{2008}{\natexlab{b}})
  \bibinfo{pages}{3524--3529}.
\bibitem[{Julien et~al.(2011)Julien, Mauger, and Zaghib}]{C0JM04190D}
\bibinfo{author}{C.~M. Julien}, \bibinfo{author}{A.~Mauger},
  \bibinfo{author}{K.~Zaghib}, \bibinfo{journal}{J. Mater. Chem.}
  \bibinfo{volume}{21} (\bibinfo{year}{2011}) \bibinfo{pages}{9955--9968}.
\bibitem[{Amin et~al.(2009)Amin, Lin, Peng, Weichert, Acartürk, Starke, and
  Maier}]{Amin_Si}
\bibinfo{author}{R.~Amin}, \bibinfo{author}{C.~Lin}, \bibinfo{author}{J.~Peng},
  \bibinfo{author}{K.~Weichert}, \bibinfo{author}{T.~Acartürk},
  \bibinfo{author}{U.~Starke}, \bibinfo{author}{J.~Maier},
  \bibinfo{journal}{Adv. Funct. Mater.} \bibinfo{volume}{19}
  (\bibinfo{year}{2009}) \bibinfo{pages}{1697--1704}.
\bibitem[{Hong et~al.(2009)Hong, Wang, Chen, Upreti, and
  Whittingham}]{hong:A33}
\bibinfo{author}{J.~Hong}, \bibinfo{author}{C.~S. Wang},
  \bibinfo{author}{X.~Chen}, \bibinfo{author}{S.~Upreti},
  \bibinfo{author}{M.~S. Whittingham}, \bibinfo{journal}{Electrochem.
  Solid-State Lett.} \bibinfo{volume}{12} (\bibinfo{year}{2009})
  \bibinfo{pages}{A33--A38}.
\bibitem[{Omenya et~al.(2011)Omenya, Chernova, Upreti, Zavalij, Nam, Yang, and
  Whittingham}]{Omenya2011}
\bibinfo{author}{F.~Omenya}, \bibinfo{author}{N.~A. Chernova},
  \bibinfo{author}{S.~Upreti}, \bibinfo{author}{P.~Y. Zavalij},
  \bibinfo{author}{K.-W. Nam}, \bibinfo{author}{X.-Q. Yang},
  \bibinfo{author}{M.~S. Whittingham}, \bibinfo{journal}{Chem. Mater.}
  \bibinfo{volume}{23} (\bibinfo{year}{2011}) \bibinfo{pages}{4733--4740}.
\bibitem[{Zhang et~al.(2011)Zhang, Liang, Ignatov, Croft, Xiong, Hung, Huang,
  Hu, Zhang, and Peng}]{Zhang2011}
\bibinfo{author}{L.-L. Zhang}, \bibinfo{author}{G.~Liang},
  \bibinfo{author}{A.~Ignatov}, \bibinfo{author}{M.~C. Croft},
  \bibinfo{author}{X.-Q. Xiong}, \bibinfo{author}{I.-M. Hung},
  \bibinfo{author}{Y.-H. Huang}, \bibinfo{author}{X.-L. Hu},
  \bibinfo{author}{W.-X. Zhang}, \bibinfo{author}{Y.-L. Peng},
  \bibinfo{journal}{J. Phys. Chem. C} \bibinfo{volume}{115}
  (\bibinfo{year}{2011}) \bibinfo{pages}{13520--13527}.
\bibitem[{Huang et~al.(2001)Huang, Yin, and Nazar}]{huang:A170}
\bibinfo{author}{H.~Huang}, \bibinfo{author}{S.-C. Yin}, \bibinfo{author}{L.~F.
  Nazar}, \bibinfo{journal}{Electrochem. Solid-State Lett.} \bibinfo{volume}{4}
  (\bibinfo{year}{2001}) \bibinfo{pages}{A170--A172}.
\bibitem[{Ellis et~al.(2007)Ellis, Kan, Makahnouk, and Nazar}]{Ellis2007}
\bibinfo{author}{B.~Ellis}, \bibinfo{author}{W.~H. Kan},
  \bibinfo{author}{W.~R.~M. Makahnouk}, \bibinfo{author}{L.~F. Nazar},
  \bibinfo{journal}{J. Mater. Chem.} \bibinfo{volume}{17}
  (\bibinfo{year}{2007}) \bibinfo{pages}{3248--3254}.
\bibitem[{Amin and Maier(2008)}]{Amin20081831}
\bibinfo{author}{R.~Amin}, \bibinfo{author}{J.~Maier}, \bibinfo{journal}{Solid
  State Ionics} \bibinfo{volume}{178} (\bibinfo{year}{2008})
  \bibinfo{pages}{1831--1836}.
\bibitem[{Islam et~al.(2005)Islam, Driscoll, Fisher, and
  Slater}]{Islam:2005p168}
\bibinfo{author}{M.~S. Islam}, \bibinfo{author}{D.~J. Driscoll},
  \bibinfo{author}{C.~A.~J. Fisher}, \bibinfo{author}{P.~R. Slater},
  \bibinfo{journal}{Chem. Mater.} \bibinfo{volume}{17} (\bibinfo{year}{2005})
  \bibinfo{pages}{5085--5092}.
\bibitem[{Fisher et~al.(2008)Fisher, Prieto, and Islam}]{Fisher:2008p80}
\bibinfo{author}{C.~A.~J. Fisher}, \bibinfo{author}{V.~M.~H. Prieto},
  \bibinfo{author}{M.~S. Islam}, \bibinfo{journal}{Chem. Mater.}
  \bibinfo{volume}{20} (\bibinfo{year}{2008}) \bibinfo{pages}{5907--5915}.

\end{thebibliography}



\end{document}